\documentclass[twocolumn,trackchanges]{aastex63}
\RequirePackage{lineno}
\usepackage{amsmath}
\usepackage{amssymb}
\usepackage{amsthm}
\usepackage{natbib,aasdefs,url,bm}
\usepackage{array}
\usepackage{float}
\usepackage{graphicx}
\usepackage{subfigure}
\usepackage{color}
\usepackage{ctable}
\usepackage{threeparttable} % to use table notes
\usepackage{CJK}
\usepackage{lineno}
\usepackage{tablefootnote}

\newcommand{\target}{MRC 0614-083}

\def\hlinewd#1{%
\noalign{\ifnum0=`}\fi\hrule \@height #1 %
\futurelet\reserved@a\@xhline} 

\shorttitle{An Accretion Flare Interpretation for the UHE Neutrino Event KM3-230213A}
\shortauthors{Yuan et al}

\begin{document}
\begin{CJK*}{UTF8}{gbsn}

\title{An Accretion Flare Interpretation for the Ultra-High-Energy Neutrino Event KM3-230213A}%

\correspondingauthor{Chengchao Yuan}
\email{chengchao.yuan@desy.de}

\author[0000-0003-0327-6136]{Chengchao Yuan (袁成超)}\affil{Deutsches Elektronen-Synchrotron DESY, 
Platanenallee 6, 15738 Zeuthen, Germany}

\author[0000-0003-2497-6836]{Leonard Pfeiffer}\affil{Julius-Maximilians-Universit{\"a}t W{\"u}rzburg, Fakult{\"a}t f{\"u}r Physik und Astronomie, Institut f{\"u}r Theoretische Physik und Astrophysik, Lehrstuhl f{\"u}r Astronomie, Emil-Fischer-Stra{\ss}e 31, 97074 W{\"u}rzburg, Germany}

\author[0000-0001-7062-0289]{Walter Winter}\affil{Deutsches Elektronen-Synchrotron DESY, 
Platanenallee 6, 15738 Zeuthen, Germany}

\author[0009-0001-9486-1252]{Jose Maria Sanchez Zaballa}
\affil{Julius-Maximilians-Universit{\"a}t W{\"u}rzburg, Fakult{\"a}t f{\"u}r Physik und Astronomie, Institut f{\"u}r Theoretische Physik und Astrophysik, Lehrstuhl f{\"u}r Astronomie, Emil-Fischer-Stra{\ss}e 31, 97074 W{\"u}rzburg, Germany}

\author[0000-0002-3308-324X]{Sara Buson}\affil{Deutsches Elektronen-Synchrotron DESY, 
Platanenallee 6, 15738 Zeuthen, Germany}
\affil{Julius-Maximilians-Universit{\"a}t W{\"u}rzburg, Fakult{\"a}t f{\"u}r Physik und Astronomie, Institut f{\"u}r Theoretische Physik und Astrophysik, Lehrstuhl f{\"u}r Astronomie, Emil-Fischer-Stra{\ss}e 31, 97074 W{\"u}rzburg, Germany}

\author[0009-0000-9401-1971]{Federico Testagrossa}\affil{Deutsches Elektronen-Synchrotron DESY, 
Platanenallee 6, 15738 Zeuthen, Germany}

%\affil{Deutsches Elektronen-Synchrotron DESY,  Platanenallee 6, 15738 Zeuthen, Germany}

\author[0000-0002-2515-1353]{Alessandra Azzollini}\affil{Julius-Maximilians-Universit{\"a}t W{\"u}rzburg, Fakult{\"a}t f{\"u}r Physik und Astronomie, Institut f{\"u}r Theoretische Physik und Astrophysik, Lehrstuhl f{\"u}r Astronomie, Emil-Fischer-Stra{\ss}e 31, 97074 W{\"u}rzburg, Germany}
%\affil{Deutsches Elektronen-Synchrotron DESY, Platanenallee 6, 15738 Zeuthen, Germany}

\begin{abstract}
We study the origin of the ultra-high-energy (UHE) neutrino event KM3-230213A detected by KM3NeT, focusing on MRC 0614-083 which has been pinpointed as the closest blazar to the neutrino localization exhibiting variable multi-wavelength emission. A joint interpretation of the optical, infrared, and X-ray light curves suggests that MRC 0614-083 has undergone a super-Eddington accretion flare accompanied by efficient proton acceleration. That flare has initiated a delayed infrared echo within the surrounding dust torus, which serves as a target for photomeson ($p\gamma$) interactions such that a self-consistent picture emerges that complements the blazar jet scenario: the predicted UHE neutrino flux is at the level expected from {joint $E^{-2}$ fit with the IceCube measurements at lower energies}, the variable nature of the event alleviates the tension with IceCube limits,
and the accompanying electromagnetic cascade describes the X-ray flare around the neutrino detection time. Since a key remaining uncertainty is the unknown redshift of the source, we strongly encourage optical/ultraviolet spectroscopic measurements to determine its redshift.

%In this study, we examine the temporal and spectral properties of the blazar MRC 0614-083, located closest to the neutrino's reconstructed position, and investigate its potential association with KM3-230213A. 
%Notably, the EM cascade scenario could reproduce the X-ray flare around the neutrino detection time. 
%Meanwhile, photomeson ($p\gamma$) interactions with infrared photons can produce UHE neutrinos at flux levels comparable to the joint $E^{-2}$ fit of current neutrino observatories, assuming a relatively low redshift (e.g., $z \sim 0.5$). 
%The transient nature of the flare may also help reconcile the absence of the detection by IceCube. 
\end{abstract}


\section{\label{sec:intro}Introduction}

The KM3NeT neutrino observatory in the Mediterranean Sea \citep{KM3Net:2016zxf} has recently reported the detection of the ultra-high-energy (UHE) neutrino event KM3-230213A in the near-horizontal direction with an energy  of 220 PeV \citep[72 PeV -- 2.6 EeV at the 90\% CL;][]{KM3NeT:2025npi}. This extraordinary event thereby extends the energy range of neutrino astronomy into the UHE domain, which means that the associated primary was an Ultra-High-Energy Cosmic Ray (UHECRs) exceeding EeV energies. 
Identifying the astrophysical source of KM3-230213A would be truly spectacular, as this may be the first direct evidence for the origin of UHECRs.
Discussed possible origins \citep{KM3NeT:2025aps, KM3NeT:2025bxl,Muzio:2025gbr,Neronov:2025jfj,Fang:2025nzg,Dzhatdoev:2025sdi,Das:2025vqd,Zhang:2025abk} include (steady or transient) extragalactic sources, Galactic sources, or a cosmogenic origin from interactions between UHECRs and background photons in the Universe. However, the non-detection of neutrinos at such energies by IceCube \citep{IceCube:2025ezc,KM3NeT:2025ccp,Li:2025tqf} results in a tension at a high confidence level ($\gtrsim3\sigma$) for the diffuse cosmogenic scenario, given IceCube's larger effective area and longer operational period. It has been also pointed out that the tension can be alleviated to $2-3\sigma$ assuming a transient point source \citep[e.g.,][]{Li:2025tqf,Neronov:2025jfj}. 
%Correlation searches and subsequent multi-messenger modeling of point sources close to the neutrino's arrival direction would be useful for uncovering its origin.
 
Accompanying the discovery of KM3-230213A, 17 blazars located within the 99\% containment region have been identified as potentially correlated with the UHE neutrino event \citep{KM3NeT:2025bxl}. Seven of these blazars have been identified through X-ray, infrared, and radio cross-matches. Three were characterized ``most promising'' by the KM3NeT collaboration because of their multi-wavelength activity \cite[see Fig. 4 in][]{KM3NeT:2025bxl}. MRC 0614-083 is the only one of these three sources located within the 90\% CL contour, thereby satisfying both selection criteria: (a) inclusion within the 90\% CL contour and (b) evidence of interesting multi-wavelength activity. This blazar is also the source closest to the most probable arrival direction of the neutrino. Interestingly, MRC 0614-083 exhibits an X-ray flare observed by the Swift X-Ray Telescope \citep[XRT;][]{SWIFT:2005ngz} and \emph{eROSITA} \citep{eROSITA:2024oyj} around the time of the neutrino detection. Additionally, optical observations from the Asteroid Terrestrial-impact Last Alert System \citep[ATLAS;][]{2018PASP..130f4505T}, the Catalina Real-Time Transient Survey \citep[CRTS;][]{2009ApJ...696..870D}, and the Zwicky Transient Facility \citep[ZTF;][]{2019PASP..131a8003M}, along with infrared observations from the Wide-field Infrared Survey Explorer \citep[WISE;][]{2010AJ....140.1868W} and Near-Earth Object WISE \citep[NEOWISE;][]{2011ApJ...731...53M}, have revealed variable features in the light curves prior to the neutrino's arrival, suggesting potential interconnections. {While MRC 0614-083 is also detected by radio surveys like the NRAO VLA Sky Survey \citep[NVSS;][]{NVSS} and Very Large Array Sky Survey \citep[VLASS;][]{VLASS} , there are no detailed, published radio studies investigating its jet opening angle or Lorentz factors.}
Motivated by the time delays between the peaks of the optical and infrared light curves, as well as the possible coincidence of the neutrino detection with the X-ray flare, we propose an accretion flare model for the blazar MRC 0614-083, {in which protons accelerated during the optical flare phase -- associated with enhanced accretion activity -- produce neutrinos and trigger electromagnetic cascades}. We aim to provide a self-consistent interpretation of the infrared and X-ray flares, while establishing a connection between the flaring state and the UHE neutrino event. In this model, the optical light curve will be treated as a proxy for the time evolution of the accretion rate onto the supermassive black hole (SMBH). Such activity may originate from enhanced accretion of ambient molecular gas. The dust torus surrounding the SMBH, with a radius of approximately $\sim10^{17}-10^{18}$ cm \citep[e.g.,][]{2017MNRAS.469..255G}, can absorb the optical emission and re-radiate it in the infrared bands, producing so-called dust echoes \citep[e.g.,][]{2016MNRAS.458..575L,2016ApJ...828L..14J}.  
%We demonstrate that, by using the observed optical light curve and assuming an appropriate dust radius, the infrared light curve can be well explained by this infrared echo mechanism. 
 
{In the context of active galactic nuclei (AGN), besides extended relativistic jets, the particle accelerators could be structures within the inner edge of the dust torus, such as accretion disks \citep[e.g.,][for review]{Murase:2022feu}, sub-relativistic winds \citep{Wang:2016oid,Liu:2017bjr}, disk coronae \citep{Murase:2019vdl,Fiorillo:2024akm}, or inner jets \citep{2012ApJ...749...63M,Murase:2014foa}. In this work, without loss of generality, we do not specify the exact acceleration mechanism, but instead inject protons into an isotropic radiation zone.} In the radiation zone, characterized by the radius of the dust torus {(such as the non-relativistic winds reaching the inner edge of the dust torus)}, $p\gamma$ interactions between injected protons and thermal optical and infrared photons can produce UHE neutrinos and initiate electromagnetic (EM) cascades \citep[e.g.,][]{Yuan:2023cmd}. The resulting neutrino emission can reach UHE ranges, and the accompanying X-ray emission may account for the observed neutrino flare. One caveat is that the redshift ($z$) of the blazar remains unknown; however, for completeness, we investigate how the outcomes of this model depend on redshift by varying $z$ from 0.5 to 2.0.

%Sec. \ref{sec:observation} describes the multi-wavelength observations and introduce the dust echo scenario used to fit the infrared light curve. In Sec. \ref{sec:model}, we compute the time-dependent neutrino and EM cascade emissions from an isotropic radiation zone in the dust torus. This section also explores the potential jet contribution and the dependence on redshift. We discuss our results and summarize our conclusions in Secs. \ref{sec:discussion} and \ref{sec:summary}.

Throughout the paper, the time and energy measured in the SMBH-rest frame are denoted as $t$ and $\varepsilon$, which are related to the quantities in the observer's frame through $t=T_{\rm obs}/(1+z)$ and $\varepsilon = (1+z)E$. We assume a standard at $\Lambda$CDM universe
with matter density $\Omega_{\rm m}=0.3$ and Hubble constant $H_0=71.9~\rm km~s^{-1}~Mpc^{-1}$ \citep{2017MNRAS.465.4914B}. The notation $Q_x$ represents $Q/10^x$ in CGS units unless otherwise specified.

\section{\label{sec:observation}multi-wavelength observations and Infrared echoes}
\subsection{Multi-wavelength observations}\label{subsec:observations}

This study combines archival multi-wavelength data with newly acquired X-ray observations to construct the spectral and temporal profile of \target. This source is detected in several radio surveys\footnote{Rapid ASKAP Continuum Survey \citep[RACS;][]{RACS}, NRAO VLA Sky Survey \citep[NVSS;][]{NVSS}, Very Large Array Sky Survey \citep[VLASS;][]{VLASSQL}, Very Large Array Low-frequency Sky Survey Redux \citep[VLLSr;][]{VLSSr}, Molonglo Reference Catalogue of Radio Sources \citep[MRC;][]{MRC}, and the Parkes-MIT-NRAO radio continuum survey \citep[PMN;][]{PMN}}, however, these observations are treated as upper limits in our model because the radio emission is likely associated with (extended) jet structures rather than the accretion-related region.

\begin{figure*}[htp]\centering
    \includegraphics[width = 0.8\textwidth]{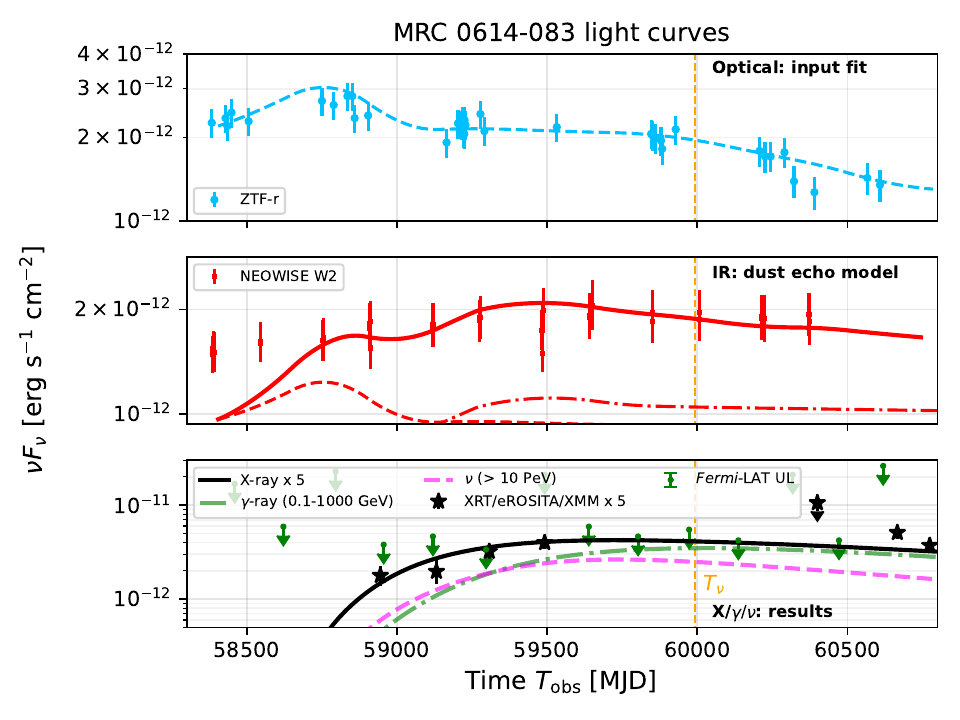}
    \caption{Observed multi-wavelength light curves of {MRC 0614-083}. Top panel: ZTF optical (r-band) observations (blue points) and the interpolated optical light curve (blue dashed curve). Middle panel: Infrared W2-band data from NEOWISE (red points) with the dust echo fit (solid red curve). The early-time and spherical components are shown as red dashed and red dash-dotted curves, respectively. Bottom panel: X-ray observations (black points), {\emph{Fermi}-LAT upper limits (green points)} and results. The cascade X-ray and $\gamma$-ray light curves are shown as the black solid and green dashed-dotted curves, respectively, while the magenta dashed curve represents the light curve of single-flavor neutrinos with energies above 10 PeV. The vertical orange dashed line depicts the neutrino detection time $T_\nu$. Data sources are listed in the main text.}
    \label{fig:LC}
\end{figure*}

In infrared (IR) and optical wavelengths we incorporate observations from WISE (W1, W2, W3 and W4 filters) and NEOWISE (W1, W2; identical to WISE) as well as ATLAS (cyan $c$ filter) and ZTF (red $r$ and green $g$ bands), covering the period from $2018$ to early $2025$. Fig.\ref{fig:LC} highlights the ZTF r-band (blue) and NEOWISE W2-band (red) light curves, shown in the first and second panels, respectively. Observations in the W3 and W4 filters are only available from the original WISE mission from $2009$ to early $2011$, since these bands were not used in later phases of the survey due to detector saturation. The magnitudes are corrected for galactic extinction and translated to energy flux units using the source-dependent reddening value between the B- and V-band, E(B-V), from \cite{Schlafly} with the extinction law adapted from \cite{Fitzpatrick}. 

The X-ray coverage includes archival data from {\textit{ROSAT}}, \textit{eROSITA} and \textit{Swift}-XRT and new Target of Opportunity (ToO) observations carried out with \textit{Swift}-XRT on the 9th of April 2025, coordinated following the neutrino detection. These data sets were uniformly reprocessed using the procedure discussed in \cite{KM3NeT:2025bxl}. {For the archival \emph{ROSAT} flux measurement, we model the spectrum as a power law with a fixed photon index of 2. Using the reported integrated flux of $1.39^{+0.47}_{-0.48}\times10^{-13}~{\rm erg~s^{-1}~cm^{-2}}$ in the 0.1$-$2.4 keV band at MJD 48150 \citep{KM3NeT:2025bxl}, we derived the corresponding power-law normalization and spectral flux density.} In the time interval MJD 58945$-$60800, the combined \emph{eROSITA} and \emph{Swift}-XRT light curve in the 0.2$-$2.3 keV energy range is shown in the bottom panel of Fig. \ref{fig:LC}. We furthermore provide a $3\sigma$ upper limit from XMM-Newton\footnote{The XMM-Newton X-ray upper limit is obtained from \url{http://xmmuls.esac.esa.int/upperlimitserver/}} at MJD 60410. 

While optical photometry and broadband spectral energy distribution (SED) data are available, no optical spectra from slit spectroscopy are publicly accessible for \target, preventing the detection of emission or absorption lines and thus a direct redshift measurement. The initial approach was to constrain the redshift photometrically using ultraviolet (UV) observations from Swift Ultraviolet/Optical Telescope (UVOT) in multiple filters, following the methods from \cite{redshiftUV}. This technique relies on the redshift-dependent onset of Lyman-alpha absorption, which causes a characteristic flux suppression in UV to optical wavelengths. Hence, we requested additional Swift observations as a target of opportunity (ToO ID: 22343) to explore this approach. The UVOT data were processed following standard procedures, using \texttt{uvotimsum} to co-add images from individual exposures and \texttt{uvotsource} to determine fluxes and magnitudes. However, the UV data only yield non-detections; we report 3$\sigma$ upper limits in each UVOT band, offering no constraint on the Lyman-alpha break and leaving ambiguity as to whether the source is intrinsically faint in the UV or fully absorbed. 
%Nevertheless, the presence of significant optical flux in the ZTF g-band (effective wavelength $\sim4747\,\textup{\AA}$) implies that the Lyman-alpha break was not shifted into this band, placing an upper limit on the redshift of $z\leq2.91$.
%OPTIONAL PART AS REDSHIFT BEEING CRUCIAL FOR MODELING:
Determining the redshift of this source is essential, as it directly impacts the luminosity scale in the joint SED fitting process. We therefore advocate for targeted spectroscopic observations at optical or UV wavelengths to obtain this key parameter.

% X-ray data reduction and spectral modeling were performed following the methodology described in \cite{Sanchez_2025}. \textit{Swift}-XRT data were processed using the \textsc{HEASoft} package (v6.33), with standard filtering and calibration applied via \texttt{xrtpipeline}. Source and background spectra were extracted from circular and annular regions, respectively, based on the point spread function of the instrument. Spectral binning was applied using \texttt{grppha} to ensure a minimum of one count per channel, allowing the use of C-statistics for low-count data. Modeling was carried out in \textsc{XSPEC} (v12.14.0h), adopting a simple absorbed power-law model that includes a fixed Galactic absorption. Data were fitted in the 0.3$-$10.0 keV.

High energy gamma-ray data was extracted from the light curve published in \cite{KM3NeT:2025bxl}, obtained with the \textit{Fermi} Large Area Telescope (LAT). The data analysis presented in the light curve yields only upper limits {(represented by the green points in the bottom panel of Fig.~\ref{fig:LC})}, with no significant flux detected around the time of the neutrino event \citep{KM3NeT:2025bxl}. Based on the provided energy range ($0.1-1000\,\text{GeV}$), we computed an upper limit on the gamma-ray energy flux at the time of the neutrino event, which serves to constrain the high energy contribution of the AGN modeling. 

To construct the final broadband SED, we averaged the fluxes from the archival and newly acquired light curves in each band and placed the resulting energy fluxes at the effective frequencies corresponding to their respective filters or pass bands. The orange, red and blue points in Fig. \ref{fig:spec} represent the radio, IR/optical/UV, and X-ray regimes, respectively, complementing the \textit{Fermi}-LAT $\gamma-$ray upper limit in green. {The archival X-ray flux level of $\sim4.4\pm2.6\times10^{-14}~\rm{erg~s^{-1}~cm^{-2}}$ in the energy range 0.1$-$2.4 keV measured at MJD 48150 \citep[][shown as the gray point in Fig. \ref{fig:spec}]{KM3NeT:2025bxl} is about 8 times lower than the X-ray measurements near the neutrino detection time, indicating that the blazar was undergoing an enhanced emission state between MJD 58945 and 60800.} The IR and optical spectra can be described as two blackbody components with temperatures of $k_B T_{\rm IR} \sim 0.1$ eV and $k_B T_{\rm Opt} \sim 1.5-2.0$ eV, respectively, in the observer's frame (see the dashed gray curve in Fig. \ref{fig:spec}). Based on these characteristic temperatures and the light curves in the $r$-band and W2-band, we estimated the bolometric luminosities using the relation $L_i = 4\pi d_L^2(z) \mathcal{C}_{i,\rm bol} [\nu F_{\nu,i}]$, where $i = \mathrm{IR}$ and $\mathrm{Opt}$ correspond to the IR and optical bands, $d_L$ is the luminosity distance, and $\mathcal{C}_{i,\rm bol}$ denotes the bolometric correction factor, defined as the ratio of the total energy flux of the blackbody distribution to the flux of the observed energy band. The bolometric IR and optical luminosities will be used in the dust echo fitting and multimessenger modeling presented in subsequent Sections, while the X-ray observations will be interpreted as the outputs of our model.

\subsection{IR: dust echoes}

{Blazar jets can produce variability; however, for MRC 0614-083, the long-term variability observed in the optical, infrared, and X-ray light curves cannot be explained solely by jet single-zone emission, as their peaks and rising phases do not occur simultaneously. {We therefore propose an accretion-flare model to self-consistently reproduce the infrared, optical, and X-ray light curves shown in Fig. \ref{fig:LC}, while a jet component is invoked to account for the far-infrared emission and the archival \emph{ROSAT} X-ray measurements.}}

Multi-wavelength observations indicate that the infrared spectrum (e.g., the first red point above 0.1 eV in Fig. \ref{fig:spec}) can be described by a blackbody distribution, in contrast to the soft component observed below 0.1 eV. The thermal shape of the spectrum, the gradually rising IR light curves, and the delayed IR emission relative to the optical emission all support a dust echo origin, similar to that seen in {candidate} neutrino-emitting tidal disruption events \citep[TDEs;][]{Reusch:2021ztx,vanVelzen:2021zsm,Li:2024qcp}, where IR photons are emitted by a dust torus heated by optical radiation from the inner regions such as the accretion disk. The temperature of the IR-emitting dust is limited by the dust sublimation temperature, $\varepsilon_{\rm IR}= k_BT_{\rm sub}\sim 0.16$ eV, implying a peak IR energy in the observer's frame of $E_{\rm IR}\lesssim3.92\varepsilon_{\rm IR}/(1+z)\simeq0.63 {~\rm eV}/(1+z)$ \footnote{The factor 3.92 corresponds to the peak of thermal emission in a $\nu F_\nu$ spectrum, obtained by solving $\frac{d}{dx}\frac{x^4}{e^{x}-1}=0.$}, consistent with the observed IR data. Within the dust echo framework, the radius of the dust torus can be estimated as $R_{\rm IR}\sim \Delta T_{\rm IR} c/[2(1+z)]\simeq 1.1\times10^{18}{~\rm cm}/(1+z)$, where $\Delta T_{\rm IR}\sim850$ d is the observed delay of the IR emission with respect to the optical peak.

To fit the IR light curve, we follow the treatment in \cite{Winter:2022fpf,Yuan:2024foi} for candidate neutrino-emitting TDEs and define the generic time-spreading function $f(t) = f_{\rm ET} + f_{\rm S}$, where $f_{\rm ET}(t) = \alpha \delta(t)$ and $f_{\rm S} = (1 - \alpha) H(t, 0, \Delta t_{\rm IR}) / \Delta t_{\rm IR}$ respectively represent the contributions from the early-time component\footnote{The early-time component could arise from an anisotropic or irregular dust distribution along the line of sight \citep{Yuan:2024foi}, or from pre-existing dust clumps around the SMBH \citep{2019ApJ...871...15J}.} and the spherical dust torus. In this expression, $0 \leq \alpha \leq 1$ denotes the fraction of the total IR power attributed to line-of-sight dust, which is modeled by the Dirac delta function $\delta(t)$. The boxcar function $H(t, 0, \Delta t_{\rm IR})$ is defined such that $H(t, 0, \Delta t_{\rm IR}) = 1$ for $0 \leq t \leq \Delta t_{\rm IR}$ and $H(t, 0, \Delta t_{\rm IR}) = 0$ otherwise, and is commonly used to describe the torus contribution. The bolometric IR luminosity $L_{\rm IR}$ is then obtained as the convolution of the bolometric optical luminosity $L_{\rm Opt}$ with the time-spreading function $f(t)$,
\begin{equation}
    L_{\rm IR}(t)=\epsilon_{\rm dust}\int_{-\infty}^{+\infty} L_{\rm opt,0}(t')f(t-t')dt',
    \label{eq:L_IR}
\end{equation}
where $\epsilon_{\rm dust}<1$ is the fraction of the incident optical radiation
that is reprocessed to IR radiation by the illuminated dust and $L_{\rm Opt,0}=L_{\rm Opt}/(1-\epsilon_{\rm dust})$ is the optical luminosity before dust absorption. Here, $f(t)$, $L_{\rm IR} (t)$, and $L_{\rm Opt,0}(t)$ are defined in terms of time ($t$) measured in the SMBH-rest frame, which is related to the time in observer's frame via $t=T_{\rm obs}/(1+z)$.

Using $L_{\rm IR}$ and $L_{\rm Opt}$ obtained before, the energy released in the IR and optical bands can be expressed as $\mathcal E_{\rm IR}=\int L_{\rm IR} dt$ and $\mathcal E_{\rm Opt}=\int L_{\rm Opt} dt$. This allows us to approximate $\epsilon_{\rm dust}$ as the ratio of the IR bolometric energy to the total energy in optical and IR bands, i.e., $\epsilon_{\rm dust}\approx \mathcal E_{\rm IR}/(\mathcal E_{\rm Opt}+\mathcal E_{\rm IR})\simeq0.37$. We fix $\epsilon_{\rm dust}=0.37$ and apply Eq. \ref{eq:L_IR} to fit the NEOWISE W2 light curve (shown as the solid red curve in the middle panel of Fig. \ref{fig:LC}), yielding $\alpha\simeq0.45$. The dashed and dashed-dotted curves respectively indicate the early-time
and spherical dust torus components, assuming a reference redshift of $z=0.5$. We restrict our fit to times after the optical peak, as IR emission before 58500 MJD likely stems from earlier optical activity and is less relevant to KM3NeT's neutrino detection window. The underestimated IR emission before 58500 MJD does not significantly affect the neutrino flux. As will be shown later, the source is optically thick to $p\gamma$ interactions. We note that the uncertainty in dust echo fit due to $z$ is not significant, as the luminosities are directly inferred from observations and scale with the luminosity distance in the same way.

\section{\label{sec:model} Multimessenger modeling}

%Meanwhile, the expected neutrino numbers at the neutrino detection time $T_{\nu}$ for KM3NeT and IceCube are respectively $\mathcal N_{\nu,\rm KM3NeT}\simeq 0.07$ and $\mathcal N_{\nu,\rm IC}\simeq 0.87.$ The tension of the non-detecti oby IceCube is around 1.91$\sigma$ (probability of $P(\rm KM3NeT-IC)=$2.8\%, Poisson.)

\subsection{Model description}
The peak of the optical light curve suggests that the blazar undergoes an accretion flare phase, which could lead to the enhanced particle acceleration. Noting that the accretion disk is likely the dominant source of AGN thermal optical emission, we assume that the accretion rate $\dot M(t)$ aligns with the optical light curve and define the accretion rate at the peak time of the optical emission, $t_{\rm pk}$, as $\dot M_{\rm pk}$. We then parametrize the luminosity of injected protons, $L_p$, as a fraction of the accretion power, i.e., $L_p=\epsilon_p \dot M c^2$, where
an efficient proton injection efficiency of $\epsilon_p=0.2$ is adopted
as a fiducial value, motivated by the studies of TDEs \citep{Winter:2022fpf}. %{Possible particle acceleration sites include the accretion disk, corona, disk wind, or inner jets. In this work, however, we do not specify the exact acceleration mechanism; instead, we assume particle injection into the isotropic radiation zone characterized by $R_{\rm IR}$ and consider subsequent interactions with isotropized target photons.}

\begin{figure*}[htp]
\centering
\includegraphics[width = 0.65\textwidth]{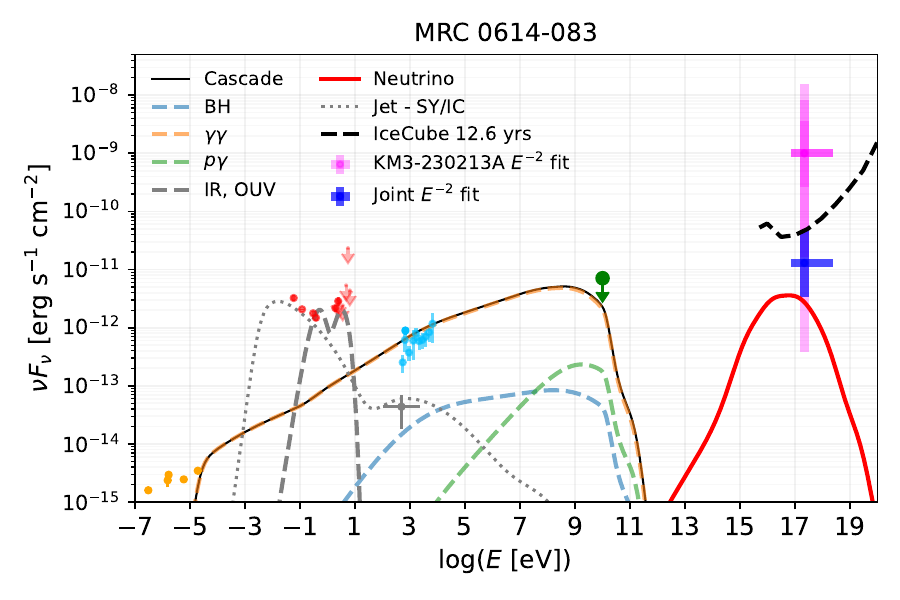}
\caption{Neutrino and cascade spectra at the neutrino detection time $T_{\nu}$ for the $z = 0.5$ case. The orange, red, blue, and green points represent the radio, infrared/optical/UV, X-ray, and $\gamma$-ray observations, respectively. The magenta point denotes the average neutrino flux derived from the KM3-230213A $E^{-2}$ fit, while the dark blue point shows the neutrino flux from the joint $E^{-2}$ fit. {The upper limit from the 12.6-year IceCube search is shown as the black dashed curve.} The solid black and red curves indicate the total cascade emission and the single-flavor neutrino flux, respectively. {The gray dashed curve represents the target photon spectrum, and the black dotted curve shows the jet interpretation to the far-infrared observations and the archival \emph{ROSAT} X-ray data (gray point)}.}
\label{fig:spec}
\end{figure*}

We assume a power-law proton injection rate, given by $\dot Q_p\propto \varepsilon_p^{-s}\exp(-\varepsilon_p/\varepsilon_{p,\rm max})$, and normalize the spectrum via $\int \varepsilon_p\dot Q_p d\varepsilon_p=L_p/V$, where $s=2$ is the spectral index, $\varepsilon_{p,\rm max}$ is the maximum proton energy and $V=4\pi R_{\rm IR}^3/3$ is the volume of the radiation zone. We assume a typical magnetic field strength of $B\sim 0.1$ G,\footnote{This value is consistent with the magnetic field strength of the disk-driven winds on sub-parsec scale in AGNs \citep[e.g.,][]{1994ApJ...434..446K,2014ApJ...780..120F}.} and will show later that the maximum proton energy $\varepsilon_{p,\rm max}$ can reach $\sim 10^9-10^{10}$ GeV.

After being injected into the radiation zone, the energetic protons interact with the isotropized thermal optical and IR photons via the $p\gamma$ process, as well as with ambient wind protons via the $pp$ process. {We also note that satisfying the $p\gamma$ threshold implies proton energies that are high enough to interact with IR target photons.} {Both the early-time and the spherical dust torus IR components could contribute to $p\gamma$ interactions.} {Using the target photon energy $\varepsilon_{\rm IR}\simeq0.16$ eV, we estimate the free-escaping optical depths at the $\Delta$-resonance threshold energy $\varepsilon_{p,\rm th}\approx (m_\Delta^2-m_p^2)c^4/(4\varepsilon_{\rm IR})\sim10^{9}~\rm GeV$} for $p\gamma$ and $pp$ processes as
\begin{equation}
    \tau_{p\gamma}\approx\frac{3\sigma_{p\gamma}L_{\rm IR}}{4\pi R_{\rm IR}c\varepsilon_{\rm IR}}\simeq1.6L_{\rm IR, 44}R_{\rm IR, 18}^{-1}\left(\frac{\varepsilon_{\rm IR}}{0.16~\rm eV}\right)^{-1}
\end{equation}
and
\begin{equation}
\begin{split}
    \tau_{pp}&\approx \frac{\eta_w\sigma_{pp}\dot M}{4\pi R_{\rm IR}v_wm_p}\\
    &\simeq 1.3\times10^{-5}R_{\rm IR, 18}^{-1}\beta_{w,-1}^{-1}\dot M_{-7}, %\left(\frac{\dot M}{10^{-7}~M_\odot~s^{-1}}\right)
    \end{split}
\label{eq:tau}
\end{equation}
where {$m_{\Delta}\simeq1.23~{\rm GeV}/c^2$ is the $\Delta$-baryon mass, $m_p$ is the proton mass,} $\sigma_{p\gamma}\simeq 500~\mu\rm b$ is the $p\gamma$ cross section, $\sigma_{pp}\simeq 40~\rm mb$ is the $pp$ cross section \citep{Kafexhiu:2014cua}, $\eta_w\sim10^{-3}-10^{-1}$ represents the fraction of accreted mass converted to the wind \citep[see e.g.,][]{2009PASJ...61L...7O,2019ApJ...880...67J}, the typical wind velocity $v_w=0.1c$ is used for wind launched from the disk \citep[e.g.,][]{2020PhRvD.102h3013Y}, and the accretion rate is written as $\dot M = 10^{-7}\dot M_{-7} ~M_\odot \rm ~ s^{-1}$. The relation $\tau_{p\gamma}\gg \tau_{pp}$ demonstrates that $p\gamma$ interactions dominate the proton cooling and pion production, as expected in radiation-dense environments. This conclusion holds even for an optimistic wind loading efficiency $\eta_w=0.1$. In the following calculations, we focus on the neutrinos and other secondary particles produced via $p\gamma$ photomeson process. 

The resulting charged ($\pi^\pm$) and neutral pions ($\pi^0$) {will} decay into neutrinos, $\gamma$-rays, and secondary electrons/positrons. These secondary particles, together with electron-positron pairs generated from $\gamma\gamma$ annihilation and Bethe-Heitler (BH) interactions, subsequently initiate EM cascade emissions via synchrotron and inverse Compton radiation. A fully time-dependent modeling of particle injection, cooling, and escape is necessary to accurately describe these nonlinear EM cascades. We use the open-source software $\rm AM^3$ \citep[Astrophysical Multi-Messenger Modeling;][]{Klinger:2023zzv} to solve the coupled transport equations for all relevant particle species. A detailed description of the transport equations can be found in \cite{Yuan:2023cmd} and \cite{Klinger:2023zzv}. {The maximum proton energy, $\varepsilon_{p,\rm max}$, is determined by the physical properties of the acceleration zone. Here, we estimate the maximum proton energy assuming that protons are initially accelerated or re-accelerated in the wind reaching to $R_{\rm IR}$,} by equating the $p\gamma$ cooling rate to the characteristic acceleration rate, given by $t_{\rm acc}^{-1} = \eta_{\rm acc} e B c / \varepsilon_p$, where $\eta_{\rm acc} \lesssim 1$ is the acceleration efficiency. {Even if the protons are not initially accelerated in the wind, they could still be energized to $\varepsilon_{p,\rm max}$ by the wind diffusive shocks and we will continue to use $\varepsilon_{p,\rm max}$ as a reference in the following discussions.}

The AM$^3$ software works with the particle density spectra $n_i=d^2N_i/dVd\ln\varepsilon_i$, where $i=p,~e,~\gamma,~\nu$ respectively represent protons, electrons/positrons, photons, and neutrinos. The photon and neutrino spectra can be converted to the flux in the observer's frame via
\begin{equation}
    \nu F_\nu(T_{\rm obs},E_i)= c\frac{R_{\rm IR}^2}{3d_L^2}\varepsilon_i n_i (\varepsilon_i)\bigg\rvert_{T_{\rm obs}/t=\varepsilon_i/E_i=1+z}.
    \label{eq:nu_F_nu}
\end{equation}
For $\gamma$-rays, an additional correction factor $\exp[-\tau_{\rm EBL}(z,E_\gamma)]$ should be multiplied to Eq. \ref{eq:nu_F_nu} to account for the $\gamma\gamma$ absorption with the extragalactic background light (EBL) during propagation through the Universe, where $\tau_{\rm EBL}$ is the optical depth as a function of redshift and photon energy.

{This work focuses on the radiation processes in the isotropic radiation zone embedded in the dust torus, however the blazar relativistic jet may also contribute to the multi-wavelength emissions. The blazar single-zone leptonic model describes high-energy emission from a relativistic jet, where a compact region of plasma containing relativistic electrons moves along the jet axis. In this model, the observed emission is primarily produced through synchrotron radiation and inverse Compton scattering, without generating high-energy neutrinos. Key parameters include bulk Lorentz factor $\Gamma$, comoving size $R'_{b}$ of the radiation zone, magnetic field strength $B'_{b}$, electron luminosity $L_e'$, spectral index $s_b$, maximum and minimum Lorentz factors, $\gamma_{e,\rm max}$ and $\gamma_{e,\rm max}$. We treat the jet contribution independently and will later show that it may account for the far-infrared and archival X-ray observations.}

\subsection{Results for the $z=0.5$ case}

We simulate the photomeson production and subsequent EM cascades in the SMBH-rest frame from $t_{\rm start} = T_{\rm start}/(1+z)$ to $t_{\rm stop}=T_{\rm stop}/(1+z)$ to interpret the X-ray light curves and spectra, and to investigate the possibility of reproducing the UHE neutrino measurements. The starting time, $T_{\rm start} = 58400 ~\rm MJD$, and stop time, $T_{\rm stop} = ~60800 ~\rm MJD$, are chosen to encompass the optical flare, neutrino detection (see the vertical dashed orange line in Fig. \ref{fig:LC}), and the latest X-ray observations. In this calculation, the peak accretion rate, $\dot M_{\rm pk}$, is treated as a free parameter, whereas fiducial values are adopted for magnetic field strength $B=0.1$ G, injection efficiency $\epsilon_p=0.2$, and the acceleration efficiency $\eta_{\rm acc}=1$. Given the redshift, the bolometric IR and X-ray luminosities, as well as the radius of the radiation zone, $R_{\rm IR}$, are inferred from observations and dust echo fitting. We define the dimensionless peak accretion rate as $\dot m_{\rm pk} = \dot M_{\rm pk}/\dot M_{\rm Edd}$, where $\dot M_{\rm Edd}$ is the Eddington accretion rate; this means that $\dot m_{\rm pk}>1$ refers to super-Eddington accretion. For the SMBH mass of $M= 10^9M_9M_\odot$ and the radiation efficiency $\eta_{\rm rad}\sim0.01-0.1$, we have $\dot M_{\rm Edd}\simeq22~M_\odot~{\rm yr^{-1}}M_9\eta_{\rm rad,-1}^{-1}$ and $L_p(t_{\rm pk})=\epsilon_p\dot m_{\rm pk}\dot M_{\rm Edd}c^2$.

As a reference, let us first consider the `nearby' blazar case at $z=0.5$. The fit to the X-ray light curve (see the bottom panel of Fig. \ref{fig:LC}) and spectrum (see Fig. \ref{fig:spec}) yields $\dot m_{\rm pk} = 1.3M_{9}^{-1}\eta_{\rm rad,-1}$ during the optical flare phase. In the quiescent phase, the accretion rate decreases to $\sim0.74M_{9}^{-1}\eta_{\rm rad,-1}$, consistent with the typical values for powerful blazars, e.g., $\dot M\sim0.1-1\dot M_{\rm Edd}$ \citep[e.g.,][]{2008MNRAS.387.1669G,2009MNRAS.399.2041G}.

The solid black and red curves in Fig. \ref{fig:spec} represent the total cascade emission and the single-flavor neutrino spectrum at the neutrino detection time $T_\nu$, respectively, obtained by fitting the X-ray spectra and light curve. The cascade emission consists of the contributions from electrons/positrons produced via $\gamma\gamma$ pair production (orange dashed), $p\gamma-$induced charged pion decays (green dashed), and the BH pair production (dashed blue). This figure demonstrates that the pair originated from the $\gamma\gamma$ channel dominates the EM cascade, as the $p\gamma$ induced secondary emission (dashed green curve) is significantly suppressed by $\gamma\gamma$ annihilation. The $\gamma$-ray upper limit obtained by $Fermi$-LAT is also satisfied, since the internal $\gamma\gamma$ annihilation depletes the secondary $\gamma$-rays above 10 GeV. The black solid, magenta dashed, and green dashed-dotted curves in the bottom panel of Fig. \ref{fig:LC} show the X-ray (0.2 - 2.3 keV), $\gamma$-ray (0.1 - 1000 GeV), and neutrino (above 10 PeV) light curves, respectively, as the outputs of the model.

\begin{figure*}[htp]
    \includegraphics[width = 0.33\textwidth]{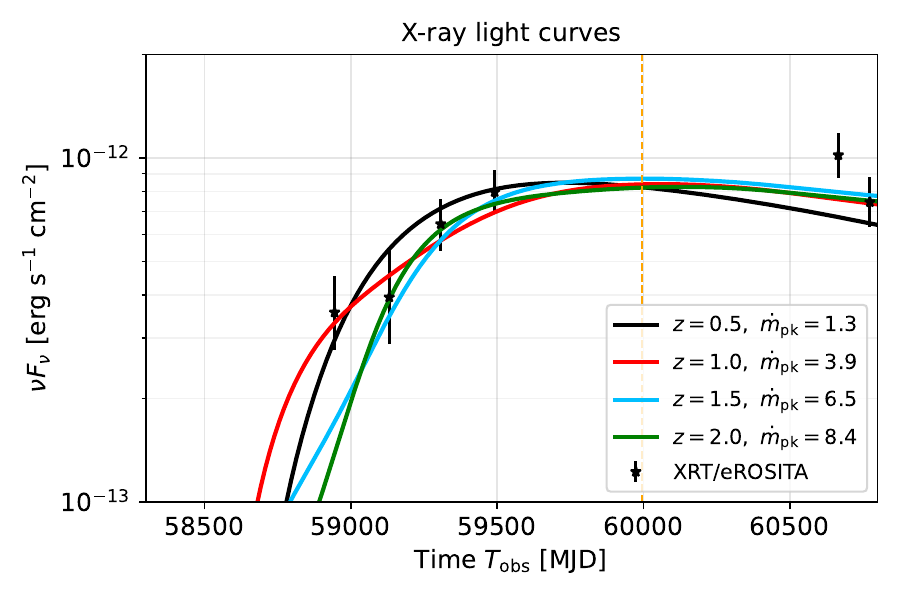}
    \includegraphics[width = 0.33\textwidth]{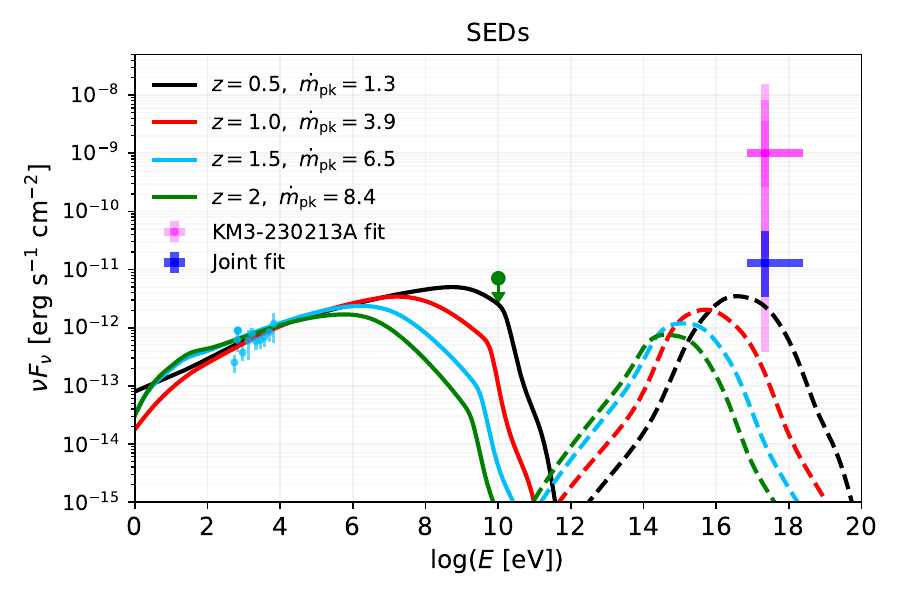}
    \includegraphics[width = 0.33\textwidth]{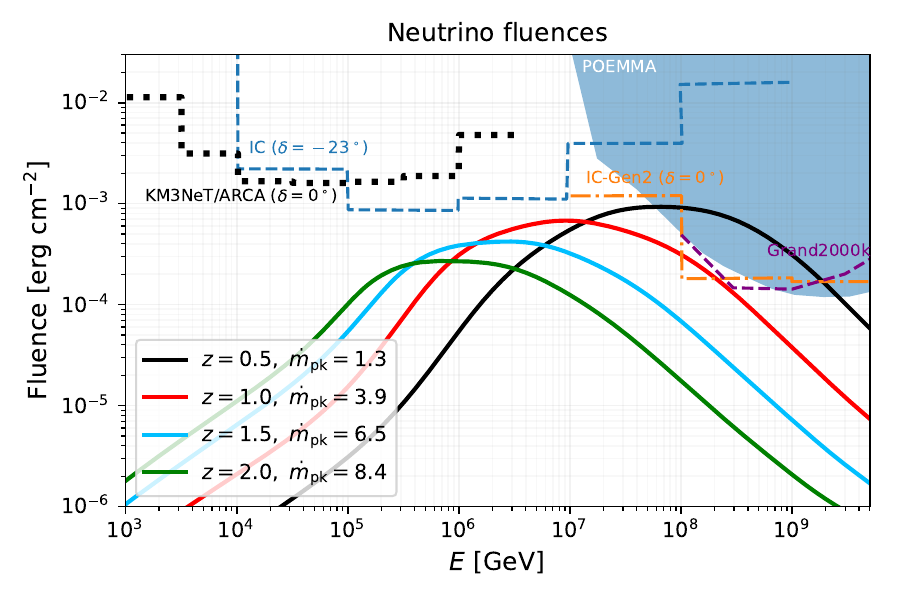}
    \caption{Impact of source redshift on model predictions, with $z$ varied from 0.5 (black curves), 1.0 (red), 1.5 (blue), to 2.0 (green). The left and middle panels show the X-ray light curve fits and spectral fits, respectively. The data points are the same as those in Fig. \ref{fig:LC} and Fig. \ref{fig:spec}. In the middle panel, solid curves represent cascade emissions, while dashed curves indicate the corresponding single-flavor neutrino fluxes. The right panel shows the cumulative single-flavor neutrino fluence up to $T_\nu$. The 90\% CL sensitivity curves of neutrino current and future detectors are also included.}
    \label{fig:test_z}
\end{figure*}

Notably, the EM cascade emission can simultaneously account for both the X-ray spectrum and the light curve in the energy range 0.2 - 2.3 keV measured by \emph{Swift}-XRT and \emph{eROSITA} (see the black solid curve in the bottom panel of Fig. \ref{fig:LC}). The time delay of the X-ray peak relative to the optical peak $\Delta T_X\approx 700-800$ d is primarily attributed to the $p\gamma$ interaction timescale $t_{p\gamma}$ which governs the development of EM cascades (see the Fig. \ref{fig:rates} in Appendix \ref{sec:appA}). From the left panel of Fig. \ref{fig:rates}, {we read that IR photons dominate $p\gamma$ interactions, which implies that $\Delta T_{X}$ is determined by the $p\gamma$ interaction time scale and the time delay of IR emissions, e.g., $\Delta T_{X}\lesssim\max[(1+z)t_{p\gamma},~\Delta T_{\rm IR}]$}. Consequently, our model naturally explains the later emerging X-ray observation together with the X-ray spectra.

For the $z=0.5$ case, the neutrino spectrum peaks at $E_\nu\simeq100$ PeV, with a flux reaching $4\times10^{-12}~\rm erg~s^{-1}~cm^{-2}$, which is close to the flux of joint $E^{-2}$ fit with the IceCube measurements at lower energies \citep[see the deep blue point in Fig. \ref{fig:spec};][]{KM3NeT:2025ccp} and lies with in the $3\sigma$ flux uncertainty of the KM3-230213A $E^{-2}$ fit \citep[magenta point;][]{KM3NeT:2025npi}. The upper limit from the 12.6-year IceCube search \citep{IceCube:2025ezc} is also shown as the black dashed curve for reference. We thus conclude that the blazar MRC 0614-083 could be a promising UHE neutrino emitter. 

We further verify that this source could accelerate protons to UHEs if the wind serves as the accelerator by evaluating the maximum proton energy $\varepsilon_{p,\rm max}$. {Here, we consider the $\varepsilon_{p,\rm max}$ that protons can reach inside the radiation zone, via wind shock acceleration (if the wind serves as the accelerator) or re-acceleration (if the protons are accelerated by other structures in the wind). {Using the $p\gamma$ interaction rates ($t_{p\gamma}^{-1}$), diffusive escape rate ($t_{\rm esc}^{-1}$) and proton acceleration rate ($t_{\rm acc}^{-1}$) presented in Appendix \ref{sec:appA}, we estimate $\varepsilon_{p,\rm max}\gtrsim10^{10}$ GeV as a good reference by balancing the proton loss rate $\kappa_{p\gamma}t_{p\gamma}^{-1}+t_{\rm esc}^{-1}$ with $t_{\rm acc}^{-1}$, where $\kappa_{p\gamma}\sim0.2$ is the inelasticity. The confinement condition, which requires the radiation zone to be larger than the Larmor radius, is naturally satisfied (see Appendix~\ref{sec:appA} for a detailed discussion).} The $\varepsilon_{p,\rm max}$ obtained here is similar to {candidate} neutrino-emitting TDEs \citep{Plotko:2024gop}. As a result, the neutrino spectrum peaks at $E_{\nu,\rm pk}\sim0.05\varepsilon_{p,\rm max}/(1+z)\gtrsim100$ PeV.} %, where $E_{\nu,0}$ is the energy that maximizes the ratio $t_{p\gamma,c}^{-1}/t_{\rm esc}^{-1}$, and $t_{\rm esc}$ represents the escape timescale of charged particles. %{While $p\gamma$ interactions between UHE protons and IR photons dominate the production of the highest-energy neutrinos, interactions with optical photons are primarily responsible for extending the neutrino spectrum down to the 100 TeV $-$ PeV range.},  similar to {candidate} neutrino-emitting TDEs \citep{Plotko:2024gop}

However, we note that the isotropic EM cascade emission predicted by the accretion flare model alone cannot fully account for the microwave and far-infrared data points in Fig.~\ref{fig:spec} (specifically, the two red points below 0.1 eV). These features are likely attributable to non-thermal emission from the relativistic jet. To illustrate this possibility, we adopt a leptonic blazar single-zone model that reproduces these low-frequency data points without delving into model details. The gray dotted curve in Fig. \ref{fig:spec} shows the synchrotron and inverse Compton emission, obtained {by fitting the far-infrared and archival X-ray data: {$\Gamma=10.8$, $R_b'=5\times10^{16}$ cm, $L_e'=5.0\times10^{43}~\rm erg~s^{-1}$, $s_b=2.3$, $\gamma_{e,\rm min}'=100$, $\gamma_{e,\rm max}'=1.4\times10^3$, and  $B'_b=4.6$ G\footnote{{This parameter set represents only one possibility, given the loose constraints on the jet parameters.}}.} }

%{The SED shown in Fig. \ref{fig:spec} is different from that of a typical blazar, such as TXS 0506+056 \citep[e.g.,][]{Keivani:2018rnh}. The synchrotron peak of MRC 0614-083 is located at a lower energy $E_{\rm sy,pk}<0.1$ eV (for reference, $\sim1-10$ eV for TXS 0506+056). This implies a relatively lower jet Doppler factor and a lower electron Lorentz factor ($\gamma_e\lesssim10-100$). In this regime, the inverse Compton (IC) spectrum is softer (steeply declining) in the X-ray range ($\sim$1-10 keV), as the IC peak energy is approximately $\gamma_e^2E_{\rm sy,pk}<1$ keV. We observe that the soft IC spectrum cannot efficiently fit the observed hard X-rays. The inclusion of hadronic components, such as proton injection, may render the jet model more plausible for fitting the X-ray data; however, the stationary one-zone jet model still has difficulties in reproducing the observed time-dependent signatures. A more complex jet structure or temporal evolution, rather than a stationary blob scenario, may be required, but this is beyond the scope of the present work.}

\subsection{Redshift dependence}

So far, we have presented a comprehensive time-domain multi-messenger analysis of MRC 0614-083, assuming a redshift of $z = 0.5$. For completeness, we now investigate the dependence of our results on redshift by varying $z$ from 0.5 to 2.0. In this calculation, the peak accretion rate $\dot{m}_{\rm pk}$ is treated as a free parameter, while $L_{\rm IR}$, $L_{\rm Opt}$, and $R_{\rm IR}$ are self-consistently determined for each redshift. All other model parameters are held fixed. The corresponding fits to the X-ray light curves and spectra are shown in the left and middle panels of Fig.~\ref{fig:test_z}. To reproduce the observed X-ray features at higher redshifts, the source must be a more powerful accretor, with $\dot{m}_{\rm pk}$ increasing from 1.3 to 8.4 as $z$ increases from 0.5 to 2.0.

The right panel of Fig.~\ref{fig:test_z} depicts the cumulative single-flavor neutrino fluence until $T_\nu$ for each redshift. The {90\% CL sensitivity curves} for various neutrino observatories, including IceCube, IceCube-Gen2 \citep{IceCube-Gen2:2021rkf}, GRAND \citep{2020SCPMA..6319501A}, POEMMA \citep{2020PhRvD.102l3013V}, and KM3NeT/ARCA230 \citep{KM3NeTARCA_sensitivity} are also shown. A notable trend is the redshift-dependent shift of the neutrino spectral peak from $\sim 100$~PeV at $z = 0.5$ to $\sim 1.5$~PeV at $z = 2.0$, consistent with expectations in the $p\gamma$ calorimetric regime, where $\kappa_{p\gamma} t_{p\gamma}^{-1} / t_{\rm esc}^{-1} \gtrsim 1$. This redshift dependence can be understood as follows: as $z$ increases, the volume of the radiation zone decreases as $R_{\rm IR}^3\propto (1+z)^{-3}$, while the IR/X-ray luminosities increase as $L_{X/\rm IR}\propto d_L^2$. Consequently, the $p\gamma$ cooling rate is enhanced as $t^{-1}_{p\gamma,c}\propto (1+z)^3d_L(z)^2$, which limits the neutrino peak energy due to the decreasing maximum proton energy, in addition to the $(1+z)^{-1}$ factor that links the energies in the source frame to the observer's frame.
%in the way: $\varepsilon_{\nu,\rm pk}\propto (1+z)^{-3}d_L(z)^{-2}$. We derive the expected neutrino peak energy in the observer's frame
%\begin{equation}
 %   E_{\nu,\rm pk}=\frac{\varepsilon_{\nu,\rm pk}}{1+z}\propto (1+z)^{-4}d_L(z)^{-2},~{\rm for}~ \frac{\kappa_{p\gamma} t_{\rm p\gamma}^{-1}}{t_{\rm esc}^{-1}}\gtrsim1.
%\end{equation}
%This analytic relation provides a reasonable explanation for the observed spectral shift. We conclude that if the source is located at higher redshift, the isotropic radiation zone becomes increasingly inefficient in producing UHE neutrinos. %In these cases, alternative zones, such as relativistic jet blobs, may play a more dominant role in neutrino production. %However, a detailed exploration of such jet scenarios is beyond the scope of this work.   

\begin{deluxetable}{c|cccc}
\label{tab:N_nu}
\tablenum{1}
\tablecaption{Expected neutrino event numbers and the tension with IceCube non-detection} %, $\eta_{\rm acc}\eta_jM_{\star}c^2\rightarrow \mathcal E_j$, data sources (e.g., SMBH mass) to be added later}
\tablewidth{0pt}
\tablehead{
{$z$} &  {$\mathcal N_{\nu,\rm KM3}$} & {$\mathcal N_{\nu,\rm IC}$}  & $P(\rm KM3\setminus IC)$  & {Tension [$\sigma$]}
}
\startdata
$0.5$ &  0.024 & 0.76 & 1.1\% & 2.30\\
 $1.0$ &  0.038  & 1.34 & 0.95\% & 2.34\\
 1.5 &  0.031 & 1.14 & 0.95\% & 2.34\\
2.0 &  0.018 & 0.72 & 0.85\% & 2.39\\
\enddata
\end{deluxetable}

\section{\label{sec:discussion} Discussion}

We have demonstrated that the accretion flare scenario can explain the time-dependent multi-wavelength observations of MRC 0614-083 and is capable of producing UHE neutrinos at a flux level consistent with the joint $E^{-2}$ spectral fit. To assess the observational prospects, it is useful to evaluate the detectability by both KM3NeT and IceCube. The UHE neutrino event KM3-230213 arrived from a near-horizontal direction to KM3NeT, corresponding to approximately 8$^\circ$ above the horizon at IceCube -- still within IceCube's optimal performance range. In this direction, the IceCube's effective area is larger than KM3NeT, which leads to a tension for the non-detection by IceCube even for transients. Using the KM3NeT horizontal effective area \citep{caiffi2024poster,Li:2025tqf}  and IceCube effective area 8$^\circ$ above the horizon \citep{IceCube:2014vjc}, we estimate the expected neutrino number ($\mathcal N_{\nu, \rm KM3/IC}$ for KM3NeT/IceCube) to be 
\begin{equation}
    \mathcal N_{\nu} = \int dE_\nu \int_{T_{\rm start}}^{T_\nu} dT A_{\rm eff}(E_\nu)F_\nu(T,E_\nu),
\end{equation}
where $A_{\rm eff}$ is the effective area. For KM3NeT, $T_{\rm start}=T_\nu - 288~\rm d$ is used noting that the exposure time is 288 days. We consider a Poissonian possibility for the KM3NeT detection with IceCube non-detection $P({\rm KM3\setminus IC})=\mathcal N_{\nu,\rm KM3}\exp({-\mathcal N_{\nu,\rm KM3}})\exp(-\mathcal N_{\nu,\rm IC})$. This probability can be translated into a confidence level for the tension, assuming a Gaussian distribution. Table \ref{tab:N_nu} summarizes the expected neutrino numbers for KM3NeT and IceCube, as well as $P({\rm KM3\setminus IC})$ and the tension confidence levels. The results indicate that for the accretion flare scenario would lead to a tension of $\sim2.3-2.4\sigma$, which is consistent with the expected values for point sources and lower compared to the tension associated with a diffuse isotropic neutrino origin or the cosmogenic scenario \citep[$3.1-3.6\sigma$;][]{Li:2025tqf}. 

%Regarding the expected number of neutrino events for IceCube, $\mathcal{N}_{\rm IC}$, discussed earlier, the integration begins at $T_{\rm start} = 58400$ MJD. In reality, however, neutrino production could have occurred during the `steady' accretion phase prior to $T_{\rm start}$. Given the $\sim 1000$-day gap without optical or X-ray observations, we estimate the maximum possible contribution from this steady phase, extending back to the start of IceCube operations. As shown in the lower panel of Fig. \ref{fig:LC}, the neutrino light curve closely follows the X-ray light curve. We therefore assume that both the X-ray and neutrino fluxes remained at a constant level prior to 58900 MJD. Under this assumption, the values of $\mathcal{N}_{\rm IC}$ listed in Table \ref{tab:N_nu} should be rescaled by a factor of $\sim 2.1$ to account for neutrino production during this extended period. This implies an increased tension with IceCube's non-detection, at the level of $2.6-2.7\sigma$, though still below the tension expected in cosmogenic scenarios. %On the other hand, the absence of X-ray observations during the steady phase may indicate a lower baryon loading, in which case the tension would converge to the values inferred from the flare phase, i.e., $2.3-2.4\sigma$.

One principal uncertainty in the accretion flare model arises from the lack of redshift measurements. The dust echo fitting and multimessenger modeling suggest that the isotropic radiation zone can account for the infrared, X-ray, and $\gamma$-ray observations by increasing only the peak accretion rate, $\dot{m}_{\rm pk}$, from 1.3 to 8.4, without altering other fiducial parameters, as the redshift increases from $z = 0.5$ to $z = 2.0$. {This required values of $\dot m_{\rm pk}$ also depend on the SMBH mass, which is not tightly constrained from current observations as well. Qualitatively, a super-Eddington accretion flare is generally needed even for the efficient proton injection case, e.g., $\epsilon_p=0.2$, and the source should be increasingly powerful if it locates at a higher redshift.} As $z$ increases, the peak neutrino energies shift from $\sim 100$ PeV to $\sim 1$ PeV, making it difficult to explain the detection of ultra-high-energy neutrinos by KM3NeT. In such a scenario, the relativistic jet with a high Lorentz factor would play an increasingly important role, motivating a lepto-hadronic modeling of the multimessenger emission from MRC 0614-083. 

%{The previous discussion indicates that the leptonic jet scenario cannot efficiently describe the SED. If hadronic injections are included, the SED can potentially be reproduced: the leptonic component accounts for the infrared spectra (similar to the dotted curve in Fig. \ref{fig:spec}), while the hadronic cascade explains the X-ray data. The limitation of this model is that the time-dependent signatures cannot be self-consistently reproduced. A more complex jet structure or evolution, rather than a stationary blob scenario, might be required. This could be an interesting avenue for future investigation but is beyond the scope of this work.}

%The EM cascade emission accounts for the X-ray observations and remains consistent with the \textit{Fermi} $\gamma$-ray upper limits throughout the duration of the flare. This conclusion is not highly sensitive to the magnetic field strength, as secondary electrons efficiently lose energy to the radiation field via synchrotron radiation and inverse Compton scattering. The high- and low-energy cutoffs of the EM spectrum are determined by in-source $\gamma\gamma$ attenuation and synchrotron self-absorption, respectively. For the neutrino spectrum, a stronger magnetic field can offset the redshift-induced energy shift. However, achieving high magnetization in an isotropic, non-relativistic region of radius $10^{17}-10^{18}$ cm is nontrivial. On the other hand, 
We have focused on the EM cascades induced by $p\gamma$ interactions, without taking into account the EM emissions from primary electrons, in order to maximize the neutrino productions. \cite{Yuan:2023cmd} pointed out that the EM cascade would dominate the multi-wavelength emission if the power of injected electrons is two orders of magnitude lower than the proton power, e.g., $L_e/L_p\lesssim10^{-2}$. This condition is generally satisfied for non-relativistic diffusive shock accelerations \citep[e.g.][]{2011JApA...32..427J,2012A&A...538A..81M}. 

\section{\label{sec:summary} Summary and conclusions}

In this work, we propose a fully time-dependent accretion flare interpretation
of the UHE neutrino event KM3-230213A in association with its directionally closest source, the AGN blazar MRC 0614-083. {The flare could originate from various processes, such as a temporary enhancement in accretion due to the infall of molecular gas.} Our generic picture is a phase of (moderately) super-Eddington mass accretion leading to the observed optical flare, followed by partially delayed emission of IR radiation (dust echoes) from re-processed radiation in  dust surrounding the SMBH. %In order to describe the emission from the dust, we use a two-component model combining dust re-emission along the site with the contribution from the dust torus.

While we cannot directly infer the proton acceleration site, the injection of non-thermal protons in an isotropic wind and their interaction with the infrared photons from the dust echoes leads to 
\begin{itemize}
    \item A neutrino signal correlated with the delayed dust echo, explaining the delay of the neutrino emission with respect to the peak mass accretion rate.
     \item An electromagnetic cascade driven by $\pi^0$ gamma-rays, which nicely describes the observed X-ray flare both as a function of time and the spectrum.
\end{itemize}

%that incorporates proton injection and the subsequent photomeson production and EM cascades in an isotropic wind region within the dust torus, aiming to interpret the multi-wavelength emissions from MRC 0614-083 and to investigate its potential association with the UHE neutrino event KM3-230213A detected by KM3NeT. 

%Our time-dependent analysis demonstrates that the accompanying IR emission can be attributed to dust echoes, while the enhanced X-ray flux observed around the time of the neutrino arrival can be well explained by EM cascade emissions associated with neutrino production. 

{Our results demonstrate that MRC 0614-083 could produce $\sim100$ PeV neutrinos during the enhanced accretion phase while simultaneously fitting the X-ray data and satisfying the \emph{Fermi}-LAT constraints.} The obtained neutrino flux is consistent with 
the level predicted by the joint $E^{-2}$ fit in combination with IceCube, provided the source is close enough (e.g., $z \lesssim 1.0$).
Moreover, given the source's variability and the fact that neutrino production reaches a high level near the time of the neutrino detection, the tension with IceCube's non-detection can be alleviated to a significance of $\sim 2.3-2.4\sigma$, regardless of the redshift. 

In conclusion, MRC 0614-083 emerges as an interesting candidate for the origin of KM3-230213A due to its proximity and its capability of producing UHE neutrinos during an accretion flare phase leading to {\em observed} dust echoes (as immediate consequence of the ``illuminated'' surrounding dust) and an {\em observed} X-ray flare (as a consequence of electromagnetic cascades associated with neutrino production), such that a self-consistent picture emerges. %The measured neutrino energies indicate the acceleration of UHECRs in the EeV range.

We recommend continued multi-wavelength follow-up observations of this source. {Optical spectroscopy can be utilized to constraint the redshift and multi-frequency radio studies can place a constraint on jet opening angles and other important jet parameters}. {A monitoring of MRC 0614-083} would enable time-domain tests of our model, facilitate future lepto-hadronic studies of potential contributions from relativistic jets, and help assess the likelihood of its association with KM3-230213A.

\acknowledgments
We thank Maria Petropoulou for valuable discussions and Georgios Vasilopoulos for providing the XMM-Newton X-ray upper limit at MJD 60410. We thank Damiano F. G. Fiorillo for insightful comments on the manuscript. This work was supported by the European Research Council, ERC Starting grant \emph{MessMapp}, S.B. Principal Investigator, under contract no. 949555. 

We acknowledge the use of public data from the Neil Gehrels Swift Observatory data archive and thank the Observation Duty Scientists and the science planners for carrying out our Swift target of opportunity observation (ToO ID: 22343).
This work has made use of data from the Asteroid Terrestrial-impact Last Alert System (ATLAS) project. The Asteroid Terrestrial-impact Last Alert System (ATLAS) project is primarily funded to search for near earth asteroids through NASA grants NN12AR55G, 80NSSC18K0284, and 80NSSC18K1575; byproducts of the NEO search include images and catalogs from the survey area. This work was partially funded by Kepler/K2 grant J1944/80NSSC19K0112 and HST GO-15889, and STFC grants ST/T000198/1 and ST/S006109/1. The ATLAS science products have been made possible through the contributions of the University of Hawaii Institute for Astronomy, the Queen’s University Belfast, the Space Telescope Science Institute, the South African Astronomical Observatory, and The Millennium Institute of Astrophysics (MAS), Chile. This work is based on observations obtained with the Samuel Oschin Telescope 48-inch and the 60-inch Telescope at the Palomar Observatory as part of the Zwicky Transient Facility project. ZTF is supported by the National Science Foundation under Grant No. AST-2034437 and a collaboration including Caltech, IPAC, the Weizmann Institute for Science, the Oskar Klein Center at Stockholm University, the University of Maryland, Deutsches Elektronen-Synchrotron and Humboldt University, the TANGO Consortium of Taiwan, the University of Wisconsin at Milwaukee, Trinity College Dublin, Lawrence Livermore National Laboratories, and IN2P3, France. Operations are conducted by COO, IPAC, and UW.
This publication makes use of data products from the Wide-field Infrared Survey Explorer, which is a joint project of the University of California, Los Angeles, and the Jet Propulsion Laboratory/California Institute of Technology, funded by the National Aeronautics and Space Administration.

%\newpage

\appendix
\begin{figure*}[h]\centering
\includegraphics[width = 0.49\textwidth]{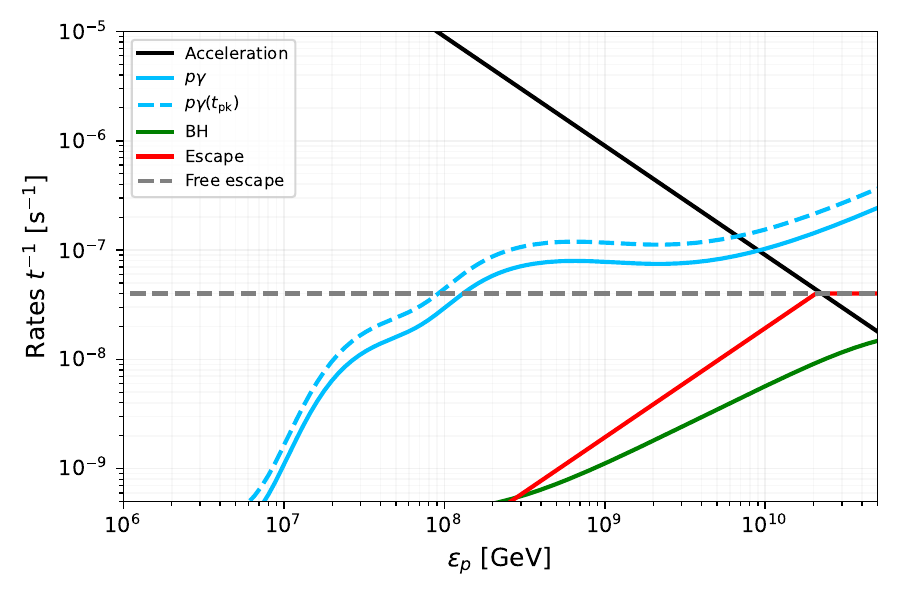}
\includegraphics[width = 0.49\textwidth]{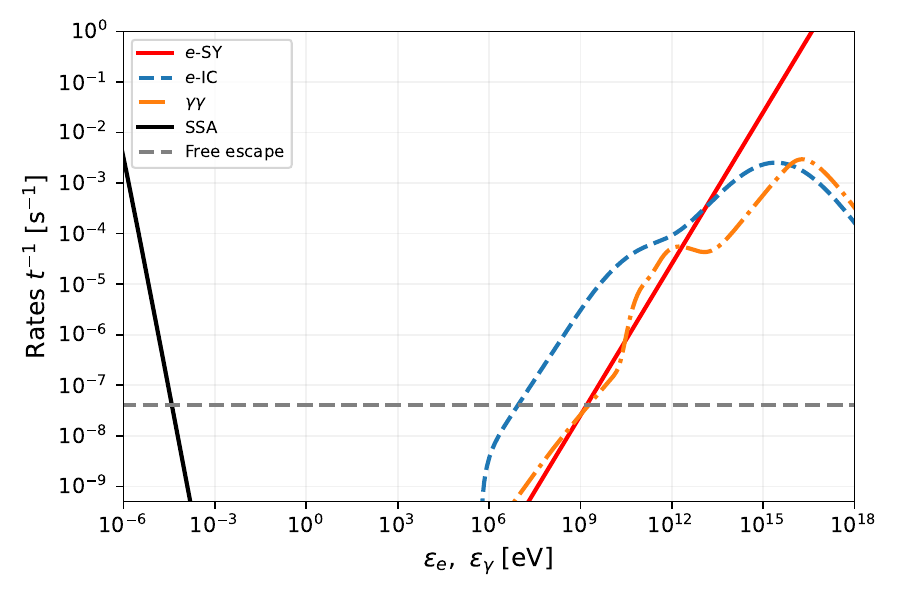}
\caption{Interaction rates at the neutrino detection time for protons (left panel), photons and secondary electrons/positrons (right panel). Physical quantities are measured in the SMBH-rest frame. Processes in the left panel include  particle acceleration (black solid), $p\gamma$ (blue), Bethe–Heitler pair production (green), particle escape (red solid). The right panel shows rates for $\gamma\gamma$ attenuation (orange dashed-dotted), synchrotron cooling (red solid), inverse Compton scattering (dark blue dashed), and synchrotron self-absorption (black solid). The horizontal gray dashed lines represent the free escape rates.}
\label{fig:rates}
\end{figure*}

\section{Interaction rates}\label{sec:appA}
In Fig. \ref{fig:rates}, we present the interaction rates for protons (left panel), photons and secondary electrons (right panel). All of the rates are obtained at the neutrino detection time except for the dashed blue curve in the left panel, which corresponds to the peak time of the optical emission $t_{\rm pk}$. From the left panel, we conclude that the acceleration of UHE protons with energies up to $\sim 10^{10}$ GeV is allowed by equating the loss rate to the acceleration rate, e.g., $\kappa_{p\gamma}t_{p\gamma}^{-1}+t_{\rm esc}^{-1}=t^{-1}_{\rm acc}$, where $\kappa_{p\gamma}\sim0.2$ is the inelasticity, $t_{\rm esc}^{-1}$ is the escape rate (red curve), %{to account for the confinement constraint}
and $t^{-1}_{\rm acc}$ is the acceleration rate with $\eta_{\rm acc}=1$ (black curve). Applying the Bohm limit, the {(diffusive)} escape rate can be written as $t_{\rm esc}^{-1}=\min[D/R_{\rm IR}^2,~t_{\rm fs}^{-1}]$ \citep{Baerwald:2013pu}, where $D=\varepsilon_p c/eB$ is the diffusion coefficient and $t_{\rm fs}^{-1}=c/R_{\rm IR}$ is the free-escaping rate. {It is constructed such that the escape reaches (as a function of energy) the free-streaming limit when the Larmor radius reaches $R_{\rm IR}$ -- which is the point where red/black solid/black dashed lines intersect in Fig. \ref{fig:rates}, left panel. Therefore, by construction, the Larmor radius must be always smaller than the size of the region in this approach.}

The electron interaction rates shown in the right panel indicate that inverse Compton scattering dominates the development of the EM cascades. In this regime, the resulting cascade spectrum follows a universal form, $\nu F_\nu \propto E_\gamma^{1/2}$ \citep{Berezinsky:2016feh}, which is consistent with the numerical results (see the EM cascade spectra in Figs. \ref{fig:spec} and \ref{fig:test_z}). 

The low-energy tail of the EM cascade emission remains consistent with the radio measurement, as the observed spectrum steepening in Fig. \ref{fig:spec} is caused by synchrotron self-absorption (SSA). The SSA break energy, $E_{\rm ssa}=\varepsilon_{\rm ssa}/(1+z)\sim2.4\times10^{-5}$ eV, where $\varepsilon_{\rm ssa}\sim3.6\times10^{-5}$ eV is determined by the condition $t^{-1}_{\rm ssa}/t^{-1}_{\rm fs}=1$ (see the right panel of Fig. \ref{fig:rates}). 

\bibliography{ref.bib}% Produces the bibliography via BibTeX.
\end{CJK*}
\end{document}